\documentclass[aps,pre,onecolumn,showpacs,superscriptaddress,groupedaddress]{revtex4}  
\usepackage{graphicx}  
\usepackage{dcolumn}   
\usepackage{bm}        
\usepackage{amssymb}   
\usepackage{amsmath}   
\usepackage{dsfont}    

\usepackage[english]{babel}
\usepackage{pgfplots}
\usepackage{textcomp}
\usepackage{color}
\usepackage{soul} 

\begin{document}

\widetext


\title{Time dependence of advection-diffusion coupling for nanoparticle ensembles}

\author{Alexandre~Vilquin}
\thanks{The authors contributed equally}
\affiliation{Gulliver CNRS UMR 7083, PSL Research University, ESPCI Paris, 10 rue Vauquelin, 75005 Paris, France}

\author{Vincent~Bertin}
\thanks{The authors contributed equally}
\affiliation{Gulliver CNRS UMR 7083, PSL Research University, ESPCI Paris, 10 rue Vauquelin, 75005 Paris, France}
\affiliation{Univ. Bordeaux, CNRS, LOMA, UMR 5798, F-33405, Talence, France}

\author{Pierre~Soulard}
\affiliation{Gulliver CNRS UMR 7083, PSL Research University, ESPCI Paris, 10 rue Vauquelin, 75005 Paris, France}

\author{Gabriel~Guyard}
\affiliation{Gulliver CNRS UMR 7083, PSL Research University, ESPCI Paris, 10 rue Vauquelin, 75005 Paris, France}
\affiliation{Universit\'e Paris-Saclay, CNRS, Laboratoire de Physique des Solides, Orsay, France}

\author{Elie~Rapha\"el}
\affiliation{Gulliver CNRS UMR 7083, PSL Research University, ESPCI Paris, 10 rue Vauquelin, 75005 Paris, France}

\author{Fr\'ed\'eric~Restagno}
\affiliation{Universit\'e Paris-Saclay, CNRS, Laboratoire de Physique des Solides, Orsay, France}

\author{Thomas~Salez}
\email[]{thomas.salez@u-bordeaux.fr}
\affiliation{Univ. Bordeaux, CNRS, LOMA, UMR 5798, F-33405, Talence, France}
\affiliation{Global Station for Soft Matter, Global Institution for Collaborative Research and Education, Hokkaido University, Sapporo, Hokkaido, Japan}

\author{Joshua~D.~McGraw}%
\email[]{joshua.mcgraw@espci.fr}
\affiliation{Gulliver CNRS UMR 7083, PSL Research University, ESPCI Paris, 10 rue Vauquelin, 75005 Paris, France}

\date{\today}

\begin{abstract}
Advection-diffusion coupling can enhance particle and solute dispersion by orders of magnitude as compared to pure diffusion, with a steady state being reached for confined flow regions such as a nanopore or blood vessel. Here, by using evanescent wave microscopy, we measure for the first time the full dynamics of Taylor dispersion, highlighting the crucial role of the initial concentration profile. We make time-dependent, nanometrically-resolved particle dispersion measurements varying nanoparticle size, velocity gradient, and viscosity in sub-micrometric near-surface flows. Such resolution permits a measure of the full dynamical approach and crossover into the steady state, revealing a family of master curves. Remarkably, our results show that the dynamics depend sensitively on the initial spatial distribution of the nanoparticles. These observations are in quantitative agreement with existing analytical models and numerical simulations performed herein. We anticipate that our study will be a first step toward observing and modelling more complex situations at the nanoscale, such as target finding and chemical reactions in nanoconfined flows, dynamical adsorption and capture problems, as well as nanoscale drug delivery systems. 
\end{abstract}

\maketitle

\section{Introduction}

\begin{figure}[t!]
	\includegraphics[scale = 1.0,trim=0 0 0 0,clip=true]{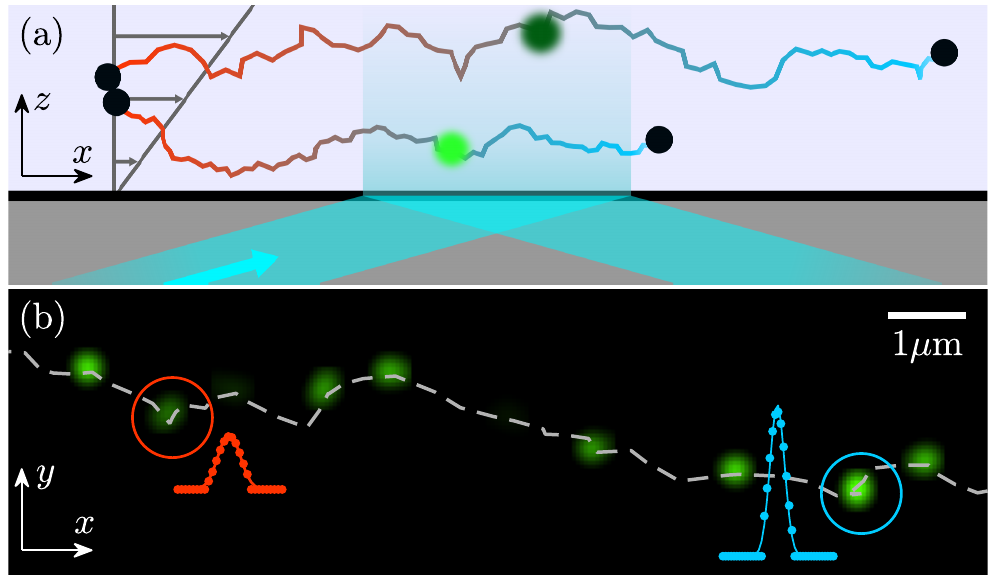}
	\caption{\label{fig:Fig1} \textbf{Taylor dispersion schematics and total internal reflection fluorescence microscopy measurements.} (a) Schematic of two Brownian colloids transported by a near surface shear flow in a microchannel illuminated by an evanescent wave. (b) Superposition of experimental images with lag time $\tau=12.5$\ ms, showing successive positions of a fluorescent 55\ nm-radius nanoparticle. Two intensity profiles are shown (arbitrary units) with the red and blue dots fitted by Gaussian profiles. The dashed gray line is a guide to the eye.}
\end{figure}

\noindent Hydrodynamic flows typically exhibit spatially varying velocity, often as a result of a nearby solid, immobile boundary. When microscopic particles are transported by such a near-surface flow, the coupling between diffusion along the flow gradients and streamwise advection leads to an enhanced dispersion as compared to the no-flow case. Indeed, the particles diffusing toward the high-velocity regions cross larger distances than the particles diffusing toward the low-velocity regions, as depicted in Fig.~\ref{fig:Fig1}(a). Such an enhancement applies to any dispersive process that couples to hydrodynamic velocity gradients, and can be orders of magnitude compared to pure diffusive dispersal. Advection-diffusion coupling is thus important in domains as diverse as pollutant spreading and drug delivery. Quantitatively, this enhancement was first described by G.I. Taylor~\cite{Taylor1953} for laminar flows inside a cylindrical tube. These predictions were applied to long times compared to the one over which a particle typically encounters the confining boundaries, here called the Taylor time. From this foundational work, and in a strict analogy to simple Fickian diffusion, a dispersion coefficient was identified as a ratio of the variance of the solute's streamwise displacement and the time. Taylor's description was formalized by Aris~\cite{Aris1956}, and generalized by Brenner and others~\cite{Brenner1993,Stone1999,Biswas2007,Griffiths2012}. This Taylor dispersion is used to measure diffusion coefficients~\cite{Bello1994, Cottet2007} and influences lab-on-chip design~\cite{Hansen2004}. 

Many theoretical and numerical works have been devoted to capture the full dynamics of Taylor dispersion, using different mathematical approaches as reviewed in a recent work~\cite{Taghizadeh2020}. Indeed, several theoretical works~\cite{Chatwin1977, VanDeVen1977, Batchelor1979, Foister1980, VanDenBroeck1982, Miyazaki1995} study the dispersion for times short compared to the Taylor time. The method of calculating different moments of the solute distribution along the flow direction, first used by Aris \cite{Aris1956} for long times, can now be used to predict the moments from short to long times \cite{Barton1983}, including the role of the initial concentration profile~\cite{Camacho1993, Vedel2012, Vedel2014}. All these works show two main results: \emph{i}) the dispersion coefficient increases from the molecular diffusion coefficient value until particles encounter the confining boundaries and then, reaches a constant long-time value predicted by Taylor; \emph{ii}) the time dependance is governed by the spatial distribution of the initial solute concentration. For single-particle observations, \emph{i.e.} an initial point concentration, the dispersion coefficient quadratically increases with time whereas an extended initial spatial distribution will lead to an additional contribution, linear in time.

Despite this intensive numerical and theoretical investigation only a few experimental studies were devoted to the short-time regime~\cite{Orihara2011, Fridjonsson2014, Takikawa2019}, and focussed only on an initial point-like concentration. Nevertheless, important applications such as drug delivery from a suddenly ruptured nanoparticle~\cite{Gentile2008,Tan2012} or predicting the efficacy of nanoconfined chemical reactions~\cite{Nielsen2010,Grommet2020} should critically depend on the initial distribution of solute~\cite{Bena2007}. Besides, Taylor dispersion plays an important role in pulmonary circulation~\cite{Fredberg1980,Grotberg1994}, cerebrospinal fluid mixing~\cite{Salerno2020}, DNA and bacterial mobility~\cite{Stein2006, Bearon2015, Dehkharghani2019} and other contexts at micro- and nano-scales~\cite{Marbach2016, Marbach2019}. In these microscopic contexts, the case of ideal, delta-like initial distributions are difficult to realise and the spatio-temporal character of the solute concentration could be critical; indeed Aminian \emph{et al.}~\cite{Aminian2016} showed the influence of the channel geometry on the Taylor dispersion and resulting concentration profiles in microfluidic chips. Therefore, experimental demonstration of the theoretical predictions describing the dispersion for all time scales and the various initial conditions is essential, but challenging. 

Among the experimental methods capable of quantitatively accessing short- and long-time dispersions at the nanoscale is total internal reflection fluorescence microscopy (TIRFM). Using this technique~\cite{Axelrod1981,Fish2009} involves illuminating a sample with an evanescent wave, which decays exponentially from the surface as shown schematically in Fig.~\ref{fig:Fig1}(a). The technique was used to investigate colloid/surface interactions, hydrodynamic boundary conditions, and hindered diffusion near a wall~\cite{Prieve1999, Pit2000, Jin2004, Yoda2011, Li2015, Yoda2020, Pit2000, Huang2006, Lasne2008, Bouzigues2008, Huang2007}. Here we use TIRFM to measure for the first time the full dynamics of Taylor dispersion for nanoparticle ensembles, as illustrated in Fig.~\ref{fig:Fig1}. We observe the predicted~\cite{VanDeVen1977, Batchelor1979, Foister1980, VanDenBroeck1982, Miyazaki1995} quadratic shear-rate dependence of the dispersion coefficient for a linear flow profile, and then study the detailed time dependence. For short-times the dispersion coefficient increases before reaching a long-time plateau. We find that a reduced form of the dispersion coefficients can be described by a family of master curves, consistent with the theoretical predictions of the moment approach~\cite{Aris1956, Barton1983, Vedel2012, Vedel2014}. As also predicted by Chatwin~\cite{Chatwin1977}, we find that the transient regime is strongly affected by the initial distribution of particles: \emph{i}) in the most general case a mixed-power time dependence is observed for the reduced dispersion coefficient; \emph{ii}) a corresponding linear time dependence can be observed for particles distributed across the whole channel; and \emph{iii}) the pure, quadratic time dependence of the dispersion coefficient for a single particle can be approached by selecting particles in the finest range of altitudes available to the experiment. These different initial conditions are accessed thanks to the nanoscale depth resolution of TIRFM.  

\vspace{2mm}

\noindent The rest of the article is structured as follows: In Section~\ref{SEC:expt} we describe the main aspects of the experimental setup and raw data analysis. In Section~\ref{sec:VelocityPDFs} we describe the velocity profiles obtained from experimental image sequences. Here, we particularly focus on the near-wall shear rate as a key parameter in the quantitative description of the Taylor dispersion. In Section~\ref{SEC:Dists} we describe measurements of the probability distributions for particle displacement. From these distributions, we study in Section~\ref{SEC:TimeGdot} diffusion and dispersion of particles as functions of time and shear rate. In Section~\ref{SEC:Theory} we turn to the theoretical descriptions of our data. These descriptions rest on the theories of Chatwin~\cite{Chatwin1977} summarised in Section~\ref{SEC:Chatwin} and Aris, Barton and Vedel \emph{et al.}~\cite{Aris1956, Barton1983, Vedel2012, Vedel2014} summarised in Section~\ref{SEC:Vedel}. After recalling the main results necessary for the theory, we investigate in Section~\ref{SEC:UniversalFamily} a family of universal experimental curves that depend on the chosen initial condition. We then conclude before presenting the appendices that are referenced throughout the text.

\section{\label{SEC:expt}Experiments}

\noindent In our experiments, fluorescent nanoparticles with radius $a = 55$ or 100\  nm were advected along the $x$-direction of a pressure-driven flow. Such advection is schematically indicated in Fig.~\ref{fig:Fig1}(a) and shown in Fig.~\ref{fig:Fig1}(b) as well as in the Supplementary Information (SI) Video 1; Appendix~\ref{sec:Methods} details the experiments. The particle volume fractions used are typically $10^{-5}$, providing sufficient statistics while avoiding hydrodynamic interactions between the nanoparticles (verification not shown). Pressure drops in the range $5\leq \Delta P\leq 400$ mbar were applied across rectangular microchannels with height $h=18$\ $\mu$m, width $w=180\ \mu$m, and length $L=8.8$\ cm. The fluid viscosities were in the range $1\leq\eta\leq7.6$\ mPa\,s, varied by controlling the concentration of glycerol/water mixtures. Given the exponentially decaying evanescent field, we call the apparent altitude of the center of mass of the particles $z=a+\Pi\ln(I_0/I)$, where $\Pi$ is the exponential decay length of \emph{ca.} 100 nm, $I$ is the measured particle intensity, as shown in Fig.~\ref{fig:Fig1}(b), and $I_0$ is the fluorescence intensity of a particle with radius $a$ at the wall, located at $z=a$. Optical aberrations lead to small deviations from an exact exponential decay for the fluorescence intensity, and such details are discussed in Appendix~\ref{sec:SIDmeanV}. Comparing subsequent images, sample trajectories as in Fig.~\ref{fig:Fig1}(b) and SI Video 2 were constructed using home-built Matlab routines. 

The experimental setup allows the observation of particles in a range of altitudes $a\lesssim z\lesssim 1$\ $\mu$m from the solid/liquid interface, the solid being a glass coverslip. Here $z=0$ is set at the solid/liquid boundary and the camera sensitivity determines the upper $z$-limit. In practice, we do not observe particles for $z\lesssim 200$\,nm as a result of electrostatic and steric interactions~\cite{Derjaguin1940, Verwey1947, Prieve1999} (see Appendix~\ref{sec:SIDmeanV} for details). The near-surface flow has the advantage of simultaneously exhibiting the lowest velocities ($0-600$ $\mu$m\ s$^{-1}$) and largest shear rates ($100-600$ s$^{-1}$) as compared to the rest of the channel, offering pertinent conditions for studying the advection-diffusion coupling. 

\section{Experimental Results}

\subsection{\label{sec:VelocityPDFs}Mean velocity profiles and shear rate measurements}

\begin{figure}[b!]
	\includegraphics[scale = 1.0,trim=0 0 0 0,clip=true]{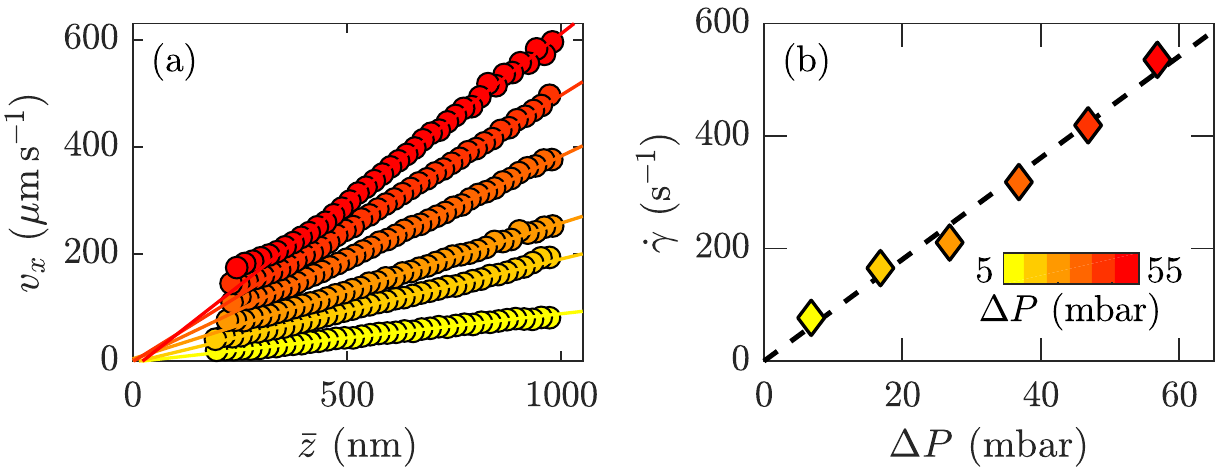}
	\caption{\label{fig:Fig2} \textbf{Velocity profiles and shear rates.} (a) Velocity profiles $v_x=\left<\Delta x\right>_\tau/\tau$ for 55\ nm-radius particles as observed in SI Videos 1 and 2 with a lag time $\tau=2.5$ ms at several pressure drops. The plain lines indicate linear regressions, providing the shear rate value $\dot{\gamma}$. (b) Shear rate as a function of pressure drop across the microfluidic channel. The dashed black line is a linear regression and the color code applies also to (a).}
\end{figure}

\noindent As schematically indicated in Fig.~\ref{fig:Fig1}(a), Taylor dispersion crucially depends on the local shear rate for particles diffusing in a shear flow. In order to obtain mean velocity profiles along the flow direction over a given lag time, $\tau$, displacements, $\Delta x(z(t),\tau)=x\left(z(t+\tau\right))-x\left(z(t)\right)$ taking $t$ as the initial observation time, were measured for each pair of frames in which identical particles were detected; Fig.~\ref{fig:Fig1}(b) and SI Video 2 show examples of particle trajectories. Since the particle intensity encodes the altitude, we first sort the particles into a series of intensity bins, each bin corresponding to a range of approximately 15 nm in the $z$-direction. Then streamwise mean velocity $v_x(\bar{z})=\langle\Delta x(\bar{z},\tau)\rangle_t/\tau$ is determined. Here, $\left<\cdot\right>_t$ denotes ensemble averaging over \emph{ca.} $10^5$ particle observations for each experimental condition accessed for all $t$. The notation $\bar{\,\cdot\,}$ denotes averaging over all frames during the lag time.

Figure~\ref{fig:Fig2}(a) shows the streamwise velocity profiles for 55\ nm-radius particles in water and several pressure drops. The solid lines show that the profiles are well approximated by linear functions (see Appendix~\ref{sec:SIDmeanV} and~\cite{Li2015,Zheng2018} for a discussion of the small non-linearities). The spread of $v_x$ intercept values arises from the spread of values for $I_0$ at each pressure, $z = 0$ corresponding to the mean of $a+\Pi\ln I_0$ over the different $\Delta P$; $z=0$ is thus resolved to within 20 nm of the solid/liquid interface assuming no slip~\cite{Li2015} as justified in Appendix~\ref{sec:SIDmeanV}. The low Reynolds numbers ($\mathrm{Re}=\rho h U/\eta\approx 10^{-2}$ with $\rho$ the fluid density and $U$ the average velocity in the whole channel) indicate a viscosity-dominated flow for which $v_x\left(\bar{z}\right)= \Delta P\left(\bar{z}^2-h\bar{z} \right)/\left(2\eta L\right)$, \emph{i.e.} a Poiseuille flow. In the region $z\lesssim 1\ \mu$m, and given the channel height $h = 18$ $\mu$m, the deviation of the Poiseuille profile from linearity is expected to be less than 5\,$\%$. Therefore, at first order in $\bar{z}/h$, we have $v_x\left(\bar{z}\right)\simeq\dot{\gamma}\bar{z}$, with the shear rate $\dot{\gamma} = \partial_{\bar{z}} v_x|_0 = h|\Delta P|/\left(2\eta L\right)$ and $\partial_{\bar{z}}$ denoting the partial derivative with respect to $\bar{z}$. In Fig.~\ref{fig:Fig2}(b) are shown the shear rate values as a function of the pressure drop extracted from velocity profiles in part (a). As highlighted by the dashed black line, the shear rate increases linearly with the pressure drop. The slope, given by $h/\left(2\eta L\right)$, provides a water viscosity $\eta=0.9\pm0.1$\ mPa\,s at 24\,\textdegree C in agreement with the expected value~\cite{Korson1969}. In Appendix~\ref{sec:ViscoDiff}, we show that $\partial_{|\Delta P|}\dot{\gamma}$ is in quantitative agreement with bulk rheological measurements of the viscosity for all the liquids investigated here.

\subsection{\label{SEC:Dists}Local distributions of particle displacement}

\begin{figure}[b!]
	\includegraphics[scale = 1.0,trim=0 0 0 0,clip=true]{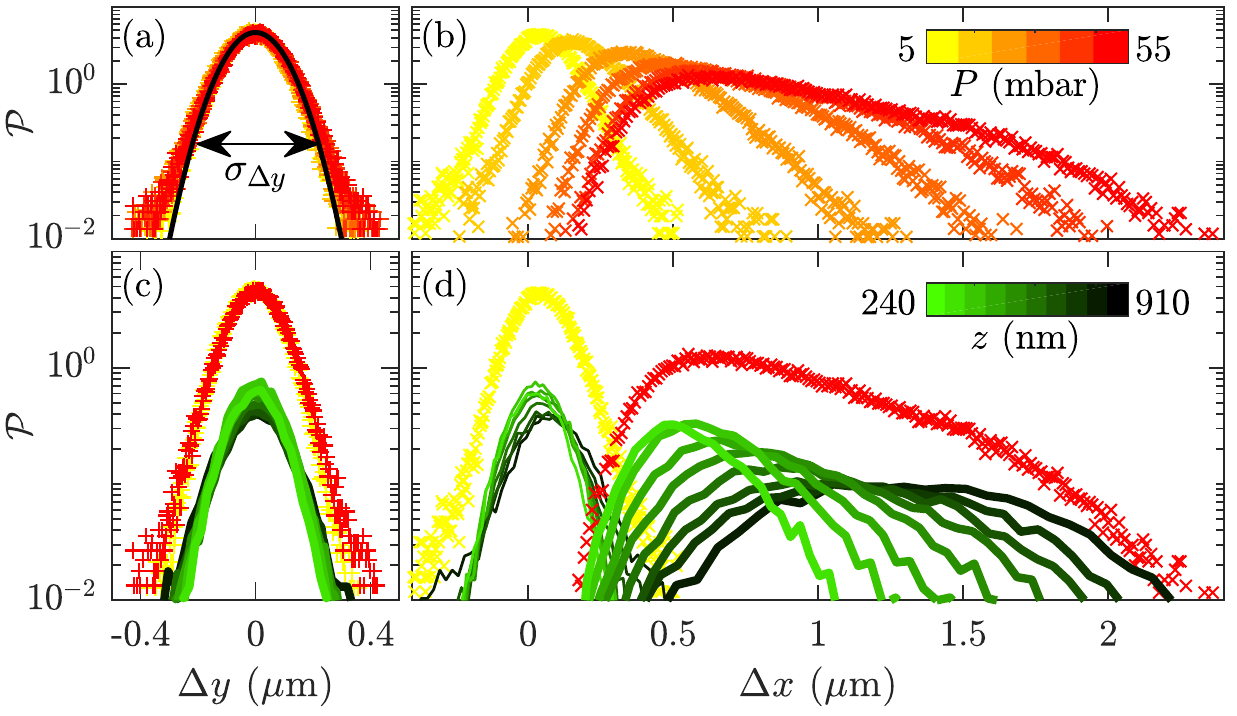}
	\caption{\label{fig:Fig3} \textbf{Probability density functions (PDFs) of particle displacements.} (a) Transverse and (b) streamwise displacement PDFs for several pressure drops indicated in the yellow-red color bar. The black line in (a) indicates a Gaussian model $\mathcal{P}\left(\Delta y\right)=\exp\left(-\Delta y^2/\left(2\sigma^2_{\Delta y}\right)\right)/\sqrt{2\pi \sigma_{\Delta y}^2}$. In (c) the $\Delta y$-PDF for the largest pressure drop is decomposed into PDFs for several $z$. Each green-black PDF corresponds to a $z$-range of \emph{ca.} 30 nm (some curves omitted for clarity), and the mean value of $z$ is indicated by the color scale of (d). The decomposed $\Delta x$-PDFs for the smallest pressure drop (thinnest green-black lines) and the largest pressure drop (thickest green-black lines), weighted by the particle number for each $z$-range in proportion to the total particle number, are shown in (d). All the displacements are measured for a time lag $\tau=2.5$ ms and concern 100 nm-radius particles in water.}
\end{figure}

\noindent Having discussed the mean velocity of the particles, we turn our attention to the displacement distributions. Figures~\ref{fig:Fig3}(a) and (b) show the probability density functions (PDFs, here called $\mathcal{P}$) of the transverse ($\Delta y$) and streamwise ($\Delta x$) displacements of 100\ nm-radius particles in water over a duration $\tau=2.5$ ms and several pressure drops. For Fig.~\ref{fig:Fig3} and in the following, we systematically take $z=z(t)$, the altitude at the initial observation time. This initial altitude $z(t)$ should be distinguished from the average altitude $\bar{z}$ over the lag time $\tau$ used in Fig.~\ref{fig:Fig2}(a). The uncertainties on the initial altitude are mainly determined by the average distance covered due to diffusion during the frame-capture time $\tau_\mathrm{capt}$, approximately equal to $\sqrt{D_0\tau_\mathrm{capt}}$ where $D_0$ is the bulk diffusion coefficient. This average distance varies from 100 nm for the 55 nm-radius particles in water to 30 nm for 100 nm-radius particles in the 50\% water-glycerol mixture. The other sources of uncertainty such as the particle-size polydispersity and the depth of field of the objective can be estimated numerically and are discussed in Appendix~\ref{sec:SIDmeanV}.

In Fig.~\ref{fig:Fig3}(a), it is shown that the transverse displacement PDFs do not depend on the pressure drop and are well described as Gaussian over two decades. The global standard deviation provides an approximation for the unidimensional Brownian diffusion coefficient $\sigma^2_{\Delta y}(\tau)/\left(2\tau\right)\approx 2.0\pm0.3$ $\mu$m$^2$\ s$^{-1}$, where $\sigma^2_{\Delta y}(\tau)=\langle\Delta y^2\rangle-\langle\Delta y\rangle^2$ is the standard deviation of the displacements along $y$ over the lag time $\tau$. This estimate is close to the value predicted by the Stokes-Einstein relation $D_0=k_\mathrm{B} \Theta/\left(6\pi\eta a\right)\approx 2.2\pm0.2$ $\mu$m$^2$\ s$^{-1}$~\cite{Einstein1905} for particles with $a=100$ nm  advected in water, where $k_\mathrm{B}$ is the Boltzmann constant, $\Theta$ is the temperature and $\eta$ was taken from bulk rheology as \emph{ca.} 0.9 mPa s at room temperature. Contrasting with the transverse displacement PDFs, those for the streamwise direction in Fig.~\ref{fig:Fig3}(b) are not Gaussian, become broadened with the pressure drop, and exhibit asymmetry as seen in Refs.~\cite{Jin2004, Aminian2016}. 

The TIRFM setup provides the particle distance from the glass/liquid interface through the detected intensity, allowing to distinguish the contributions of particles at different altitudes to the global PDFs. In the transverse direction, the local PDFs (green shades) are shown for the largest pressure drop in Fig.~\ref{fig:Fig3}(c) and they are all Gaussian regardless of $z$. In Fig.~\ref{fig:Fig3}(d) are shown similar decompositions for the smallest and largest pressure drops for streamwise displacements. These decompositions demonstrate that the asymmetry of the global distributions is mainly due to the superposition of different mean displacements at different altitudes. For the smallest pressure drop (yellow) the local PDFs are only slightly shifted with increasing $z$ due to the relatively low mean velocities, \emph{cf.} Fig.~\ref{fig:Fig2}(c). For the largest pressure drop (red), the mean values are shifted more strongly with increasing $z$ as a result of the higher shear rate. More importantly, the local PDFs thus provide access to the transverse diffusion along $y$ and streamwise dispersion along $x$ for different altitudes.

\subsection{\label{SEC:TimeGdot}Time and shear-rate dependence of the reduced dispersion}

\begin{figure}[b!]
	\centering
	\includegraphics[scale = 1.0,trim=0 0 0 0,clip=true]{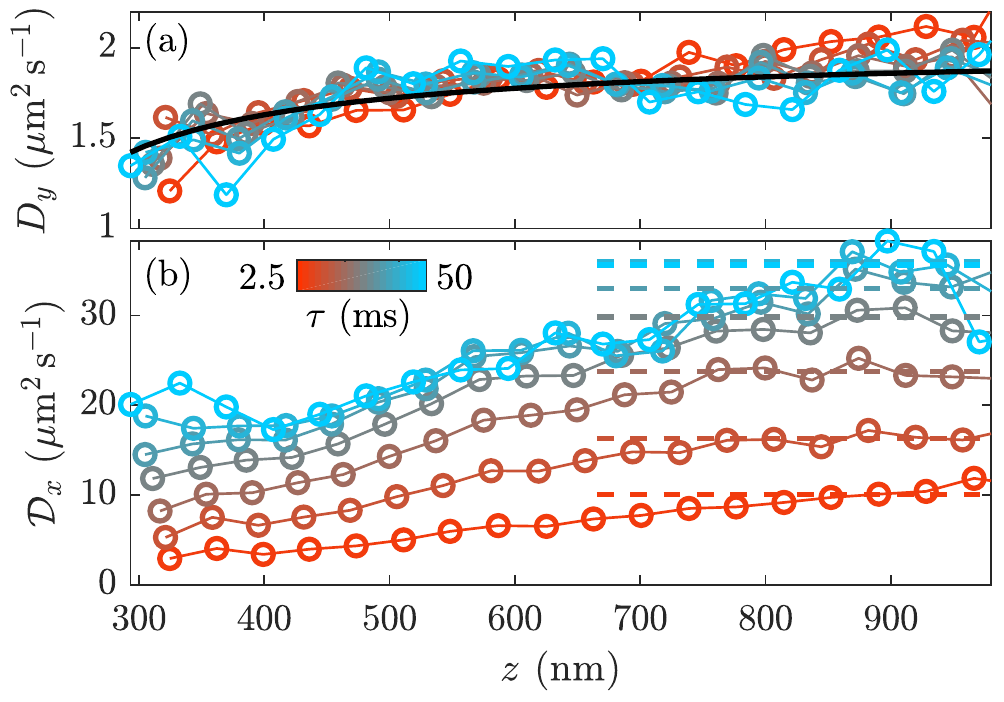}
	\caption{\label{fig:Fig4}\textbf{Spatio-temporal diffusion and dispersion coefficients.} (a) Transverse diffusion coefficients, $D_y$, and (b) streamwise dispersion coefficient, $\mathcal{D}_x$, as a function the apparent altitude and lag time, $\tau$, for 100 nm-radius particles in pure water, for a pressure drop corresponding to a shear rate $\dot{\gamma}=243$ s$^{-1}$. The $z$-range is typically 30 nm for each initial altitude $z$. In (a), the plain black line corresponds to the the theoretical prediction $D_y = D_0\left(1-\left(9/16\right)Z^{-1}+\left(3/8\right)Z^{-3}-\left(45/256\right)Z^{-4}-\left(1/16\right)Z^{-5}\right)$ with $Z=z/a$~\cite{Brenner1961}. In (b) the dashed lines indicate the streamwise dispersion coefficient, $\overline{\mathcal{D}}_x$ shown in Fig.~\ref{fig:Fig5}(a) as a function of lag time for smaller shear rates.}
\end{figure}

\noindent  A detailed study of the local Taylor dispersion as function of time and the various physical parameters at stake is now described. In Figs.~\ref{fig:Fig4}(a) and (b) are shown local transverse diffusion coefficients, 
\begin{equation}
D_y(z,\tau) = \frac{\sigma^2_{\Delta y(z)}}{2\tau}\ , 
\end{equation}
and streamwise dispersion coefficients, 
\begin{equation}
\mathcal{D}_x(z,\tau) = \frac{\sigma^2_{\Delta x(z)}}{2\tau}\ . 
\end{equation}
The latter comprise pure diffusive and advection effects, and are remarkably larger (up to an order of magnitude) than the former. These data were obtained from altitude decompositions as in Fig.~\ref{fig:Fig3} for several lag times, $\tau$ (red to blue). Figures~\ref{fig:Fig4}(a) and (b) show that there is a general increase with $z$ of $D_y$ and $\mathcal{D}_x$ until reaching a plateau at large $z$; in Appendix~\ref{sec:ViscoDiff}, we show that the plateau values are in quantitative agreement with the Stokes-Einstein relation for all liquids investigated after invoking independently measured bulk viscosities. The variation with $z$ in both directions is due to hydrodynamic interactions between particles and the solid/liquid interface, leading to a hindered diffusion as discussed in detail elsewhere~\cite{Brenner1961,Faucheux1994,Huang2007,Saugey2005}. Indeed, the $z$-dependence of $D_y$ is in agreement with the prediction resulting from the effective viscosity near a flat, rigid wall~\cite{Brenner1961} (see the plain black line in Fig.~\ref{fig:Fig4}(a)). As expected, the transverse diffusion is not dependent on the lag time $\tau$; in contrast, the dispersion coefficients, $\mathcal{D}_x$, increase significantly with $\tau$ as shown in Fig.~\ref{fig:Fig4}(b).

\begin{figure}[b!]
	\centering
	\includegraphics[scale = 1.0,trim=0 0 0 0,clip=true]{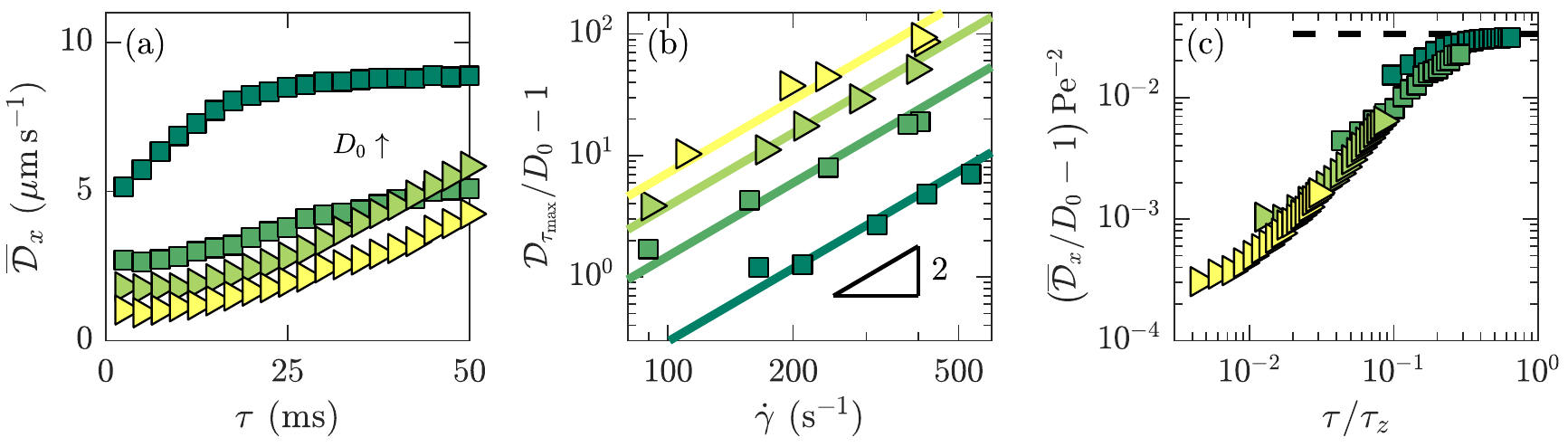}
	\caption{\label{fig:Fig5}\textbf{Shear rate and time dependances of the dispersion coefficients.} (a) Streamwise dispersion coefficient $\overline{\mathcal{D}}_x$ extracted from the dashed lines as for the example shown in Fig.~\ref{fig:Fig4}(b) as a function of lag time. From yellow to dark green, the bulk diffusion coefficient increases. In particular ---from bottom to top--- the shear rates, particle radii and viscosities are $\{\dot{\gamma}, a, \eta\} = \{110, 100, 7.6\},\,\{91, 100, 2.1\},\,\{90, 100, 1\},\,\{165, 55, 1\}$ in units \{s$^{-1}$, nm, mPa\,s\}; giving $D_0 = \{$0.28, 0.89, 2.1 3.9\}\,$\mu$m$^2$\,s$^{-1}$ determined from the Stokes-Einstein relation; marker shape indicates the liquid as water ($\square$) or glycerol-water mixtures ($\triangleright$). (b) Reduced late-time dispersion coefficient versus the shear rate; the solid lines have log-log slope $2$. (c) Same data as in (a), with time scaled using $\tau_z = H^2/D_0$ for each liquid on the $x$-axis, and the reduced dispersion coefficient scaled with $\mathrm{Pe}^2$ according to Eq.\eqref{eq:PecletNumber} on the $y$-axis; the dashed line shows the long-time limit with a prefactor predicted by Eq.~\eqref{eq:TAglobalShear}.}
\end{figure}

In Fig.~\ref{fig:Fig5}(a) we show the lag-time dependence of the streamwise dispersion coefficient. To do so, we define $\overline{\mathcal{D}}_x\left(\tau\right)$ as the average value of $\mathcal{D}_x(z(t),\tau)$ in the large-$z$ plateau for $z(t)\gtrsim H/2$ where $H$ is the size of the observation zone, typically 1~$\mu$m. That is, $\overline{\mathcal{D}}_x\left(\tau\right)$ is a conditional averaging for particles beginning their trajectory in the top half of the observation zone, thus limiting the aforementioned lubrication effects. In this figure, the bulk diffusion coefficient $D_0$ was varied by changing the particle size and the liquid viscosity. For the lowest values of $D_0$, $\overline{\mathcal{D}}_x$ continuously increases with time and the temporal slope increases; by contrast, $\overline{\mathcal{D}}_x$ saturates to a plateau for the largest $D_0$. As explained by Taylor~\cite{Taylor1953}, the time needed to reach the dispersion plateau corresponds approximately to the time needed to diffuse across the channel height. Here the Taylor time is taken as 
\begin{equation}
\tau_z = \frac{H^2}{D_0}\ . 
\end{equation}
In a rectangular channel, the exact calculation~\cite{Taylor1953} gives a characteristic diffusion time $\tau_z / \pi^2$. For the 55 nm-radius particles in water, assuming a length scale $H\approx 700$\ nm, $\tau_z/\pi^2 \approx 13$\ ms, in reasonable agreement with the corresponding data of Fig.~\ref{fig:Fig5}(a), with $D_0 = 3.9\pm0.4$ $\mu$m$^2$\ s$^{-1}$. For smaller values of $D_0$, the dispersion remains mainly in the short-time, increasing-slope regime. Nevertheless, taking the longest-time data (the data of Fig.~\ref{fig:Fig5}(a) at $\tau = 50$ ms, denoted $\mathcal{D}_{\tau_\mathrm{max}}$) for each $D_0$ value, we now examine the shear-rate dependence of the dispersion.

In Fig.~\ref{fig:Fig5}(b) is shown the dependence of the reduced late-time dispersion coefficient, $\mathcal{D}_{\tau_\mathrm{max}}/D_0-1$, for four $D_0$ as a function of the shear rate. The solid lines (with slope 2 in log-log representation) show that the reduced $\mathcal{D}_{\tau_\mathrm{max}}$ increases quadratically with the shear rate $\dot{\gamma}$ for all $D_0$ studied. To understand this result, we use the classical Taylor-Aris calculation for a plane Couette flow \cite{Brenner1993} in a rectangular channel, giving
\begin{equation}
\mathcal{D}_x=D_0\left(1+\frac{1}{30}\mathrm{Pe}^2\right) \quad \tau \gg \tau_z\ ,
\label{eq:TAglobalShear}
\end{equation}
for the infinite-time dispersion coefficient, identifying the P\'eclet number as 
\begin{equation}
\mathrm{Pe}=\frac{\dot{\gamma}H^2}{2D_0} = \frac{\dot{\gamma}\tau_z}{2}\ ,
\label{eq:PecletNumber}
\end{equation}
where we have identified the mean velocity of the flow as $u = \dot{\gamma}H/2$. Equations~\eqref{eq:TAglobalShear} and~\eqref{eq:PecletNumber} highlight the key role of velocity gradients in enhanced dispersion, and justify the quadratic shear-rate dependance of the data in Fig.~\ref{fig:Fig5}(b) for the 55 nm-radius particles in water. Applying Eq.~\eqref{eq:TAglobalShear} to this latter data, we extract a length scale $H\approx 500$\ nm, consistent with the range of $z$ observed in Fig.~\ref{fig:Fig2}(a). Because the lower-$D_0$ data does not reach the infinite-time plateau, the prefactor for the linear regressions does not reveal the corresponding size of the flow region $H$. However, as we show in the following, the quadratic shear rate dependence is preserved for all time regimes, explaining the scaling of all the data in Fig.~\ref{fig:Fig5}(b).

We now examine the detailed time dependence of dispersion coefficient for all of the experimentally accessed times. Partly inspired by previous theoretical works~\cite{Barton1983, Vedel2012, Vedel2014, Taghizadeh2020} and by the shear-rate dependence of Fig.~\ref{fig:Fig5}(b), we show in Fig.~\ref{fig:Fig5}(c) the reduced $\overline{\mathcal{D}}_x$ normalized by $\mathrm{Pe}^2$ as a function of the dimensionless lag time, $\tau/\tau_z$. Remarkably, the data in Fig.~\ref{fig:Fig5}(a) collapse onto a single master curve, suggesting the existence of a universal function describing the reduced dispersion coefficient. 

In Fig.~\ref{fig:Fig5}, we considered particles beginning their trajectories in the top half of the channel. However, this fraction can be generalized, with $n$ representing the fraction of the observation zone from which particles leave, and with bounds $0\leq n\leq1$. The reduced dispersion is thus expected to follow a relation of the form
\begin{equation}
\left(\frac{\mathcal{D}_{\langle n\rangle}}{D_0}-1\right)\mathrm{Pe}^{-2}=\mathcal{F}_{\langle n\rangle}\left(\frac{\tau}{\tau_z}\right)\ .
\label{ReducedDispCon}
\end{equation}
Examining Fig.~\ref{fig:Fig5}(c), we note that the reduced $\overline{\mathcal{D}}_x$ increases with time until reaching a plateau. According to Eq.~\eqref{eq:TAglobalShear}, in the $\tau/\tau_z\rightarrow\infty$ limit, $\mathcal{F}_{\langle n\rangle}$ should reach $1/30$. This value is shown with the horizontal black dashed line and is in quantitative agreement with the late-time data. We note furthermore that the crossover to this late-time regime occurs when $\tau/\tau_z \approx 1$. Concerning the data at the shortest dimensionless lag times in Fig.~\ref{fig:Fig5}(c), we find that they do not follow the typically predicted~\cite{VanDeVen1977, Batchelor1979, Foister1980, VanDenBroeck1982, Miyazaki1995} and measured~\cite{Orihara2011,Fridjonsson2014,Takikawa2019} early-time $\tau^2$ dependence. A key foundation of this $\tau^2$ dependence is the assumption that each particle begins its trajectory at the same initial altitude. In general, however, particles may leave from a non-peaked distribution of initial altitudes. This distribution is particularly relevant for Fig.~\ref{fig:Fig5}(c), since the plotted quantity is related to an average of particles leaving mainly from the top half of the observation zone (indicated by the dashed lines in Fig.~\ref{fig:Fig4}(b)). 

\section{\label{SEC:Theory}Theory}

\noindent In the case of a non-peaked initial distribution, Chatwin~\cite{Chatwin1977} theoretically predicted that dispersion coefficients are modified by an additional term linear in time affecting the short-time regime. After, Barton, Vedel \emph{et al.}~\cite{Barton1983, Vedel2012, Vedel2014} produced a moment theory for all times that verify this initial-concentration dependence and show plateaus at long times when particles encounter the boundaries. For both theories, the particles are considered as tracers, neglecting the electrostatic and hydrodynamic interactions with the walls~\cite{BRENNER1977312}. In the following two subsections, we outline the results from these theories that allow to recover the experimental observations we have described in the previous sections, justifying Eq.~\eqref{ReducedDispCon}. 

\subsection{\label{SEC:Chatwin}Role of the initial concentration in the short-time regime}

\noindent We first focus on the short-time regime of the dispersion and the effect of the initial conditions. While several works predicted~\cite{VanDeVen1977, Batchelor1979, Foister1980, VanDenBroeck1982, Miyazaki1995} and measured~\cite{Orihara2011,Fridjonsson2014,Takikawa2019} a quadratic time dependance for a sharply peaked initial spatial distribution, Chatwin~\cite{Chatwin1977}, recalling Saffmann~\cite{Saffman1960}, showed that for particles having an initial PDF $\mathcal{P}_i\left(y_i,z_i\right)$ and advected with a steady flow velocity $v_x\left(y_i,z_i\right)$, the dispersion coefficient at short times is given by
\begin{multline}
\frac{\mathcal{D}_{x}}{D_{0}}-1 = \frac{1}{2}\frac{\tau}{D_0}\left[\iint v_{x}^{2}\mathcal{P}_i\mathrm{d}y_{i}\text{d}z_{i}-\left(\iint v_{x}\mathcal{P}_i\mathrm{d}y_{i}\mathrm{d}z_{i}\right)^{2}\right] + \\ \frac{1}{6}\tau^{2}\left[\iint\left[\nabla^{2}v_{x}^{2}+v_{x}\nabla^{2}v_{x}\right]\mathcal{P}_i\mathrm{d}y_{i}\mathrm{d}z_{i}-3\left(\iint v_{x}\mathcal{P}_i\mathrm{d}y_{i}\mathrm{d}z_{i}\right)\left(\iint v_{x}\nabla^{2}\mathcal{P}_i\mathrm{d}y_{i}\mathrm{d}z_{i}\right)\right],
\label{eq:DeffChatwin}
\end{multline}
where $y_i$ and $z_i$ are the coordinates for the initial distribution. Assuming a 2D problem in the $Oxz$ plane with an initial particle distribution $\mathcal{P}_{z,i}=\int \mathcal{P}_i\mathrm{d}y_{i}$ and a linear shear flow $v_{x}\left(z\right)=\dot{\gamma}z$ as justified by our experiments, Eq.~\eqref{eq:DeffChatwin} simplifies to
\begin{equation}
\frac{\mathcal{D}_{x}}{D_{0}}-1=\dot{\gamma}^{2}\tau\frac{\left\langle z_{i}^{2}\right\rangle -\left\langle z_{i}\right\rangle ^{2}}{2D_{0}}+\frac{1}{3}\dot{\gamma}^{2}\tau^{2}\left[1-\frac{3}{2}\left\langle z_{i}\right\rangle \left(\int z_{i}\partial_{z_{i}}^{2}\mathcal{P}_{z,i}\text{d}z_{i}\right)\right]\ .
\label{eq:DeffChatwinLinear}
\end{equation}
This form for the dispersion clearly shows a quadratic shear-rate dependence, as observed in Fig.~\ref{fig:Fig5}(b). Furthermore, Eq.~\eqref{eq:DeffChatwinLinear} shows that the linear temporal term is weighted by the variance of the initial spatial distribution of the particles. 

Next, we make a the further simplifying assumption that particles are uniformly distributed at the initial time. The particles are thus distributed over a vertical segment of length $nH$ and centered at altitude $\left<z_i\right>$. The initial PDF is therefore:
\begin{equation}
\mathcal{P}_{z,i}\left(z_i\right) = \left\{
    \begin{array}{ll}
        0 & \mbox{if } \left|z_i-\left<z_i\right>\right|>nH/2 \\
        1/nH & \mbox{if } \left|z_i-\left<z_i\right>\right|\leqslant nH/2
    \end{array}
\right.
.
\label{eq:ProbaZ0}
\end{equation}
Injecting this distribution into Eq.~\eqref{eq:DeffChatwinLinear}, using the P\'eclet number of Eq.~\eqref{eq:PecletNumber} and with $\tau_z=H^2/D_0$, the reduced dispersion coefficient of Eq.~\eqref{ReducedDispCon} for short times becomes
\begin{equation}
\mathcal{F}_{\langle n \rangle}\left(\frac{\tau}{\tau_z}\right) = \frac{n^2 }{6}\left(\frac{\tau}{\tau_z}\right)+\frac{4}{3}\left(\frac{\tau}{\tau_z}\right)^2, \quad \tau \ll \tau_z\ .
\label{eq:TAshortShear}
\end{equation}
This relation is valid for all $n$, and in particular, we denote the $n=\{0,1/2,1\}$ cases as ``dot'', ``half-line'', and ``line'' conditions. Before discussing the results of this early-time theory, we first describe a general theory for all times.

\subsection{\label{SEC:Vedel}Full dynamics using the moment theory}

\noindent Given the asymptotic behaviors for both short and long times, we now use the framework based on concentration moments of Aris, Barton and Vedel \emph{et al.}~\cite{Aris1956, Barton1983, Vedel2012, Vedel2014} to obtain the dispersion coefficient at all times for a linear shear flow. We consider the 2D problem of tracers with an initial concentration $c_i\left(z\right)$ advected along $x$ with a mean velocity $U=\left<v_x\right>_z$. We nondimensionalize the variables through $C=c/\left<c_i\right>_z$, $X=x/H$, $Z=z/H$, $T=tD_0/H^2$, $V\left(Z\right)=v_x\left(z\right)/U$ and the P\'eclet number $\mathrm{Pe}=UH/D_0$. The 2D advection-diffusion equation therefore becomes:
\begin{equation}
\partial_{T} C+\mathrm{Pe}V(Z)\partial_X C=\partial^2_X C+\partial^2_Z C,
\label{eq:AdvecNoDim}
\end{equation}
with the initial conditions $C(X,Z,0) = C_i(Z)\delta(X)$ and $\delta$ the Dirac delta function. The $p^\mathrm{th}$ concentration moment $C_p\left(Z,T\right)$ and the corresponding average moment $M_p\left(T\right)$ are defined as:
\begin{equation}
 C_p(Z,T) = \int_\mathbb{R} X^p C(X,Z,T) \, \mathrm{d}X \quad \textrm{and} \quad M_p(T) = \int_0^1 C_p(Z,T) \, \mathrm{d}Z.
\end{equation}
These moment definitions can be introduced into the advection-diffusion equation after multiplication by $X^p$ and integration. Then, assuming no flux at the boundaries and that $\lim_{X \rightarrow \pm \infty}X^\mu \partial^\nu_X C = 0$ for arbitrary positive integers $\mu$, $\nu$, the following recursive equations are obtained: 
\begin{align}
 \left(\partial_{T} - \partial_Z^2\right) C_p(Z,T) & = p(p-1) C_{p-2} + \mathrm{Pe} \,V(Z)\,p \, C_{p-1}\ , \\
 \partial_{T}  M_p(Z,T) & = p(p-1) \langle 1,C_{p-2}\rangle_Z + \mathrm{Pe} \,p \, \langle V,C_{p-1}\rangle_Z\ ,
\end{align}
where the notation $\langle F,G\rangle_Z$ stands for the usual scalar product $\langle F,G\rangle_Z = \int_0^1 F(Z)G(Z)\mathrm{d}Z$. These equations can be recursively solved to get the first moments and then, the dimensionless dispersion coefficient. For our purposes, this latter is defined as $\mathcal{D}_x/D_0=\left(M_2-M_1^2\right)/\left(2T\right)$. This definition is used to be consistent with the Chatwin theory described in Section~\ref{SEC:Chatwin} and our experiments, even while the definition used in the Vedel \emph{et al.} work~\cite{Vedel2012,Vedel2014} is $\mathcal{D}_x/D_0=\left(1/2\right)\partial_T\left(M_2-M_1^2\right)$. 

While the details of solving for the moments are found in Refs.~\cite{Barton1983,Vedel2012,Vedel2014}, the calculations lead to a general form for the dispersion coefficient expressed as an infinite series of orthonormal functions $f_k(Z)$ and their associated eigenvalues $\lambda_k$ satisfying $(\partial_Z^2+\lambda_k)C=0$ with no flux at the boundaries:
\begin{multline}
\left(\frac{\mathcal{D}_x}{D_0} - 1\right)\mathrm{Pe}^{-2} = T^{-1}\sum_{k,j \in\mathbb{N}^2} \bigg[\bigg(\gamma_{1,k} \alpha_{0,k} T+\alpha_{1,k} \bigg)\delta_{j,k} + \alpha_{0,k} \beta_{j,k} \bigg] \left\langle V,f_{j}\right\rangle_Z g_{1}\left(k,T\right) -\gamma_{1,k}\alpha_{0,k}\delta_{j,k}\left\langle V,f_{j}\right\rangle_Z g_{2}\left(k,T\right) \\
-\frac{1}{2}\alpha_{0,k}\alpha_{0,j}\left\langle V,f_{k}\right\rangle_Z \left\langle V,f_{j}\right\rangle_Z g_{1}\left(k,T\right)g_{1}\left(j,T\right)\ .
\label{eq:TAallShear}
\end{multline}
Here, $f_{k\geq 1}(Z) = \sqrt{2} \, \cos(k\pi Z)$, $\lambda_k = k^2 \pi^2$, $f_{0}(Z)$ = 1 and $\lambda_0 = 0$. The coefficients $\alpha_{j,k}$, $\beta_{j,k}$ and $\gamma_{j,k}$ with $(j,k)$ in $\mathbb{N}^2$ are defined as
\begin{align}
 \alpha_{1,k} & = -\sum_{j \in \mathbb{N}} \alpha_{0,j} \beta_{k,j}\ , & \beta_{j,k} & = (1 - \delta_{j,k})\frac{\langle f_j,V,f_k\rangle_Z}{\lambda_j - \lambda_k}\ , & \gamma_{1,k} & = \langle f_k,V,f_k\rangle_Z\ ,
\label{eq:TAallShearCoefs}
\end{align}
with the initial conditions contained in the coefficient $\alpha_{0,j} = \langle f_j,C_i\rangle_Z$. The functions $g_{1}\left(k,T\right)$ and $g_{2}\left(k,T\right)$ are defined as
\begin{equation}
g_{1}\left(k,T\right)=\begin{cases}
\begin{array}{c}
T\\
\frac{1-\exp\left(-\lambda_{k}T\right)}{\lambda_{k}}
\end{array} & \begin{array}{c}
\mathrm{if\ }k=0\\
\mathrm{if\ }k>0
\end{array} \ ,
\qquad \mathrm{and} \qquad
g_{2}\left(k,T\right)=\begin{cases}
\begin{array}{c}
\frac{1}{2}T^{2}\\
\frac{T}{\lambda_{k}}+\frac{\exp\left(-\lambda_{k}T\right)-1}{\lambda_{k}^{2}}
\end{array} & \begin{array}{c}
\mathrm{if\ }k=0\\
\mathrm{if\ }k>0
\end{array}\end{cases}\end{cases}.
\end{equation}
The solution given by Eq.~\eqref{eq:TAallShear} is valid for any flow profile and any initial concentration profile in one dimension. 

To calculate the particular solutions for a linear shear flow $V\left(Z\right)=2Z$ and uniform distributions defined in Eq.~\eqref{eq:ProbaZ0} for the dot, half-line, and line conditions, we follow Barton \cite{Barton1983}. Injecting the linear shear flow $V\left(Z\right)=2Z$ into Eq.~\eqref{eq:TAallShearCoefs}, we first find:
\begin{equation}
\gamma_{1,0} = 1\ , \quad \mathrm{and} \quad \gamma_{1,k\geq 1}  = \int_0^1 2\cos^2(\pi k Z)\,2Z\, \mathrm{d}Z  = 1\ ,
\label{EQ:Gamma}
\end{equation}

\begin{equation}
\psi_{0}  = \langle V,f_0\rangle_Z = \int_0^1 \,2Z\, \mathrm{d}Z = 1\ , \quad \mathrm{and} \quad \psi_{k\geq1}  = \langle V,f_k\rangle_Z = \int_0^1 \sqrt{2}\cos(\pi k Z)\,2Z\, \mathrm{d}Z = -\frac{2 \sqrt{2} \bigg(1 - (-1)^k\bigg)}{\pi ^2 k^2}\ ,
\end{equation}

\begin{equation}
\beta_{0,k}  = \frac{\int_0^1 \sqrt{2}\cos(\pi k Z)\,2Z\, \mathrm{d}Z}{\lambda_0 - \lambda_k}  =-\frac{\psi_{k}}{\lambda_k} = -\beta_{k,0}\ ,
\label{EQ:Beta}
\end{equation}

\begin{equation}
\beta_{j,k} = \frac{\int_0^1 2\cos(\pi k Z)\cos(\pi j Z)\,2Z\, \mathrm{d}Z}{\lambda_j - \lambda_k}  = -\frac{2}{\pi^2(j^2-k^2)}\, \bigg[ \frac{1 - (-1)^{j-k}}{\pi^2(j-k)^2} + \frac{1 - (-1)^{j+k}}{\pi^2(j+k)^2} \bigg], \quad \quad \text{ if } j,k \geq 1\ .
\end{equation}
Then, we can evaluate the coefficients $\alpha_{0,k}$ and $\alpha_{1,k}$ that depend on the initial concentration field $C_i\left(Z\right)$. Considering uniform distributions we find 
\begin{align}
\alpha_{0,0}=1\ ,\qquad\qquad\alpha_{0,k>0}=\sqrt{2}\cos\left(k\pi \langle Z_i\rangle\right)\mathrm{sinc}\left(\frac{k\pi n}{2}\right),\qquad\text{and}\qquad\ \alpha_{1,k} = -\sum_{j \in \mathbb{N}}a_{0,j}\beta_{k,j}\ ,
\label{EQ:alpha}
\end{align}
where $\langle Z_i\rangle$ and $n$ are defined as in Eq.~(\ref{eq:ProbaZ0}) and $\mathrm{sinc}(x) = \sin(x)/x$. Inserting Eqs.~(\ref{EQ:Gamma}), (\ref{EQ:Beta}), and (\ref{EQ:alpha}) in Eq.~\eqref{eq:TAallShear} allows to generate predictions for the dispersion coefficients at all times.

\begin{figure}[t!]
	\includegraphics[scale = 1.0,trim=0 0 0 0,clip=true]{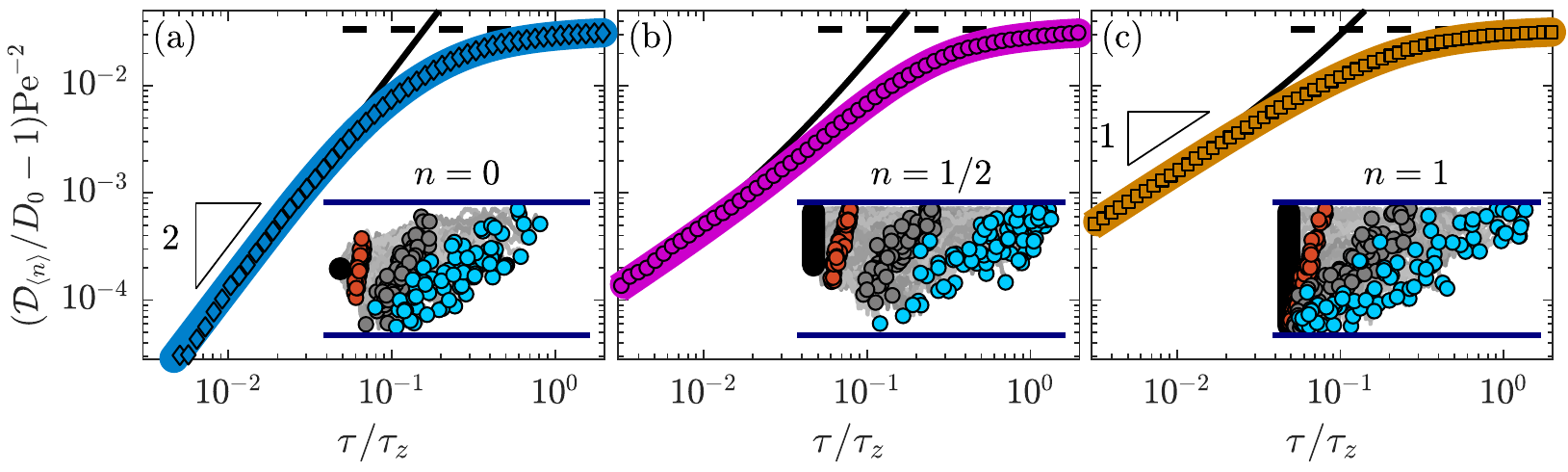}
	\caption{\label{fig:Fig6} \textbf{Dispersion from theory and simulations.} Reduced dispersion coefficients as a function of dimensionless time obtained from the Langevin simulations, Eqs.~\eqref{eq:XLangevin} and~\eqref{eq:ZLangevin} studied for fractions (a) $n=0$, (b) $n=1/2$, and (c) $n=1$ of the channel height. The black dashed and solid lines respectively correspond to the asymptotic behaviors for the long- and short-time regimes predicted by Eqs.~\eqref{eq:TAglobalShear} and~\eqref{eq:TAshortShear}. The solid colored lines show the dispersion coefficients predicted by Eq.~\eqref{eq:TAallShear} from the moment theory for the corresponding initial conditions. The insets schematically show three instants of particle trajectories advected from the associated observation zone in a linear shear flow and with diffusion in $z$. The slope triangles in (a) and (c) respectively denote power-law exponents 2 and 1.}
\end{figure}

\subsection{Langevin simulations}

\noindent In addition to making use of the Chatwin and moment theories, we performed numerical integrations of overdamped Langevin equations assuming tracer-like particles where gravity, electrostatic and hydrodynamic interactions with the walls are neglected and are instead replaced by reflective boundary conditions at the top and bottom of the channel of height $H$. The discrete coupled Langevin equations read:
\begin{align}
x\left(t+\delta t\right) & = x\left(t\right) + \dot{\gamma}z\left(t\right)\delta t+\sqrt{2D_0\delta t} \, S\left(0,1\right)\ , \label{eq:XLangevin} \\
z\left(t+\delta t\right) & = z\left(t\right) + \sqrt{2D_0\delta t} \, S\left(0,1\right)\ , \label{eq:ZLangevin}
\end{align}
where $S(0,1)$ is a Gaussian noise with zero mean and unit standard deviation, and $\delta$t is the time step. Using the same nondimensionalization as in the previous section, we integrate these equations with the results described in detail in the next section. The insets in each panel of Fig.~\ref{fig:Fig6} illustrate the Langevin simulations with initial conditions (black), trajectories (grey lines) and snapshots (red to blue circles) for several particles, along with the reflecting boundaries (horizontal blue lines); see also SI Video 3. 

\subsection{Theoretical results}

\noindent Figure~\ref{fig:Fig6} summarises the theoretical and computational results described above. First, symbols show the reduced dispersion coefficients obtained from the Langevin simulations for the (a) dot, (b) half-line and (c) line conditions. These results show an initial increase with time, and a plateau reached at long times. This observation is consistent with the general trends observed in the experimental results of Fig.~\ref{fig:Fig5}(c). Horizontal, black dashed lines show the asymptotic long-time limit for a linear shear flow, the prefactor value of 1/30 shown in Eq.~\eqref{eq:TAglobalShear}, also consistent with the experimental data. 

The solid black lines show our implementations of the Chatwin theory expressed by Eq.~(\ref{eq:TAshortShear}) for the values of $n$ indicated in the insets. A null variance, the dot condition, of the initial distribution arises if all particles start at the same altitude. For this condition, the classical $\tau^2$ dependence~\cite{VanDeVen1977, Batchelor1979, Foister1980, VanDenBroeck1982, Miyazaki1995, Orihara2011,Fridjonsson2014,Takikawa2019} for the reduced short-time dispersion coefficient is recovered. This quadratic dependence reflects a steadily increasing diversity of newly sampled velocities (see SI Videos 3 and 4) contributing to the enhanced dispersion. For non-vanishing initial variance, the early-time reduced dispersion coefficient has a linear temporal evolution. This linear behavior for extended distributions results from particles at different altitudes transported over different distances by the linear shear flow (SI Videos 3 and 4).  At times longer than the crossover time $\tau_\mathrm{C}=3(\langle z_i^2 \rangle - \langle z_i \rangle^2)/(2D_0)$ ---obtained by setting the linear and quadratic terms of Eq.~\eqref{eq:TAshortShear} to be equal--- a quadratic time dependence is recovered. When the initial variance is small with respect to $H^2$, this crossover occurs before the Taylor plateau is reached. 

Lastly, colored solid lines show the full dynamic theory of Vedel and coworkers, expressed in Eq.~\eqref{eq:TAallShear}. For all initial conditions depicted in Fig.~\ref{fig:Fig6}, the Vedel solution captures the full dynamics of the dispersion accessed by the Langevin simulations and is consistent with the two asymptotic behaviors described by the long-time Taylor description and the short-time Chatwin theory.  While the moment theory has the advantage to describe the dispersion for all time scales, we note that the asymptotic solution based on Chatwin's work~\cite{Chatwin1977} provides a simpler picture concerning the effect of the initial conditions in the short-time regime.

\section{\label{SEC:UniversalFamily}Master curves for time-dependent Taylor dispersion}

\begin{figure}[t!]
	\includegraphics[scale = 1.0,trim=0 0 0 0,clip=true]{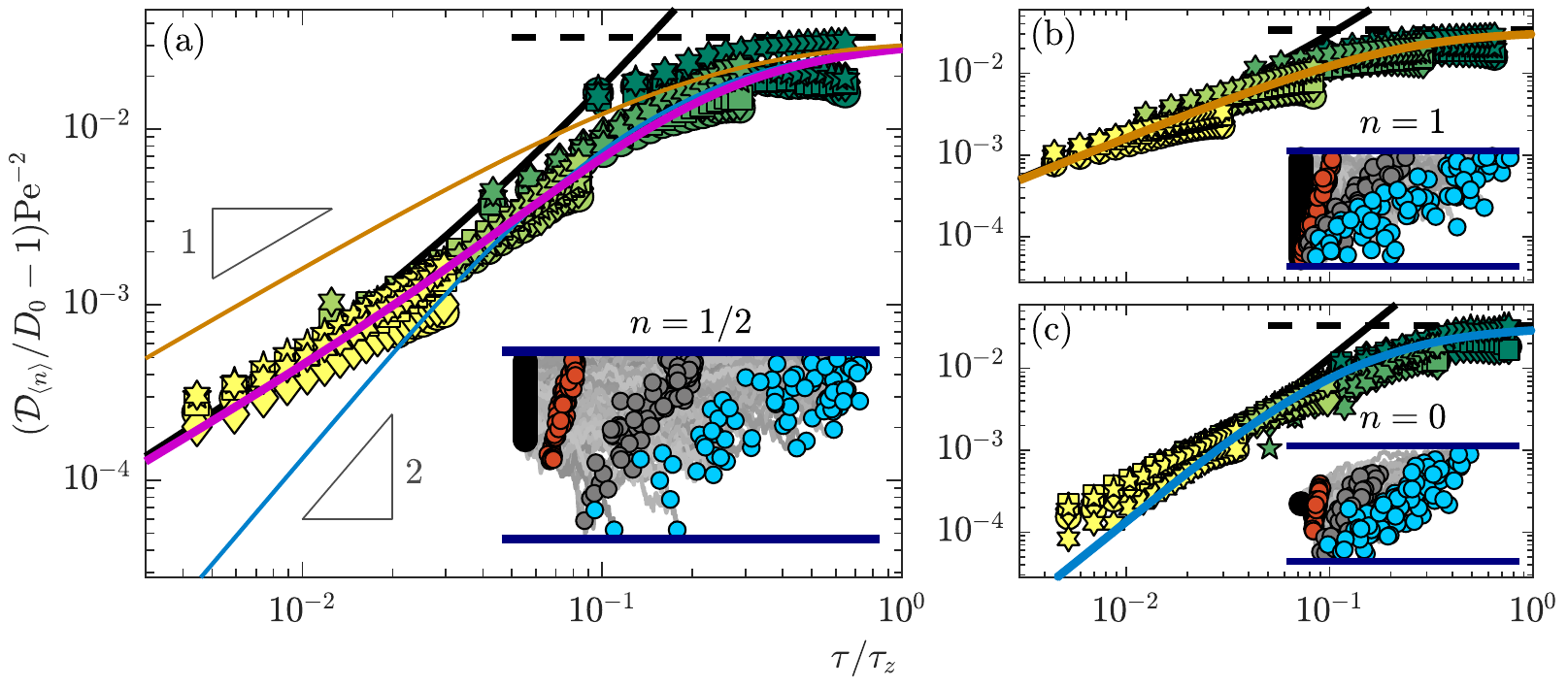}
	\caption{\label{fig:Fig7} \textbf{Dispersion from theory and experiment.} Reduced dispersion coefficients as a function of dimensionless time for all shear rates and $D_0$ studied for fractions (a) $n=1/2$, (b) $n=1$, and (c) $n=0$ of the observation zone. The bulk diffusion coefficients, $D_0$, for the data from dark green to yellow are identical to those in Figs.~\ref{fig:Fig3}(c) and (d). The black dashed and solid lines respectively correspond to the asymptotic behaviors for the long- and short-time regimes predicted by Eqs.~\eqref{eq:TAglobalShear} and~\eqref{eq:TAshortShear}. The colored wide lines show the dispersion coefficients predicted by Eq.~\eqref{eq:TAallShear} from the moment theory for the $n=1/2$, $n=1$ and $n=0$ conditions. Part (a) contains the three cases to highlight the different scaling laws whereas Figures (b) and (c) only contain the corresponding solution. The insets schematically show three instants of particle trajectories advected from the associated observation zone in a linear shear flow and with diffusion in $z$. The slope triangles in (a) denote power-law exponents 1 and 2.}
\end{figure}

\noindent Having reported the theoretical predictions in the particular case of a linear shear flow, we now compare with our experimental measurements. In Fig.~\ref{fig:Fig7}(a), we show the reduced dispersion $\overline{\mathcal{D}}_x$ normalized by the P\'eclet number $\mathrm{Pe}^2$ as a function of the dimensionless lag time, $\tau/\tau_z$. Remarkably, the data of Fig.~\ref{fig:Fig5}(c), along with that for experiments implementing four other shear rates per $D_0$, still collapse onto a single master curve. As can be seen, there is a remarkable agreement between the experimentally measured dispersion coefficient and the moment theory given by Eq.~\eqref{eq:TAallShear} for the half-line condition, with $n=1/2$. As explained in Section~\ref{SEC:TimeGdot}, the average dispersion $\overline{\mathcal{D}}_x$ is built with small $z$-ranges of 30 nm for a total $z$-range of approximatey 700 nm. Thus, it is perhaps surprising to find an agreement between a nominal value of $n=30/700\approx 0.043$ and the half-line condition. However, we recall that a range of \emph{apparent} altitudes defined by $z=a+\Pi\ln(I_0/I)$ is selected, which are not precisely the \emph{real} altitudes. As detailed in Appendix~\ref{sec:SIDmeanV}, the particle intensities are functions of the altitudes, depth of field and position of the objective, and particle sizes. With such ingredients, we can show numerically that a fine distribution of apparent altitudes actually corresponds to a larger and non-uniform distribution of real altitudes. For the case selected here, $n=1/2$ gives the best description of the data, in particular showing the power-law transition at early times and the long-time saturation. 

In order to test the analytical models, we again leverage the depth resolution of the TIRFM method to select different initial distributions. First, we choose the line condition illustrated in the inset of Fig.~\ref{fig:Fig7}(b). We thus study the dispersion coefficients for all observed particles, meaning that we consider the global distributions of Fig.~\ref{fig:Fig3}(b). The reduced dispersions measured for this condition shown in Fig.~\ref{fig:Fig7}(b), are systematically above the previous half-line condition shown in Fig.~\ref{fig:Fig7}(a). This confirms the theoretical prediction that the short-time regime is modified because particles initially leave from an extended region of space. Moreover, the results are in quantitative agreement with the full-time moment theory, and in particular demonstrates the early-time, linear-power predicted by Chatwin. Similarly, we consider particles leaving from a narrow altitude range, approaching the dot condition (inset of Fig.~\ref{fig:Fig7}(c)). In Fig.~\ref{fig:Fig7}(c) is shown the corresponding temporal evolution of the reduced dispersion coefficient for particles leaving from $z_i/H=0.21\pm0.02$. While the data does not reach exactly the theoretical prediction at early times for the dot condition due to the polydispersity of the particles (and thus non-unique altitudes for a given intensity~\cite{Zheng2018} as described above for the half-line condition), it is systematically below the one for the half-line condition and approaches the $\tau^2$ asymptotic behavior predicted by Eq.~\eqref{eq:TAshortShear} and the moment theory with $n=0$ in Eq.~\eqref{EQ:alpha}. The data analysis and analytical theory used for three different initial particle distributions using the same measurement data demonstrate the crucial role of such an initial particle distribution for the short-time dispersion.

\section*{CONCLUSION}

\noindent We report on an experimental, theoretical, and numerical study of advection-enhanced dispersion, providing the first experimental validation of analytical models predicting the full time dynamics of Taylor dispersion. Our study particularly highlights the crucial role of the initial condition in the short time regime. First, we experimentally demonstrate that the two regimes share the same shear-rate dependence, the shear rate being particularly large for near-surface transport. Furthermore, we reveal and characterize how the initial particle distribution affects the short-time dispersion. Specifically, we observe for the first time, the short-time, mixed-power-law behavior for the general case before reaching a crossover to the well-known long-time saturation regime for linear shear flows. In the extremal cases of \emph{i}) full-channel observations, a linear approach in time of the dispersion coefficient to the long-time value is observed, while for \emph{ii}) fine near-surface resolutions a quadratic temporal tendency is approached. Altogether, the experimental data are in quantitative agreement with the analytical predictions and results from Langevin numerical simulations. In the rich context of particle transport, such concepts should prove pertinent in quantitative prediction and observation of time-dependent, near-surface nanoparticle and solute dispersion, with applications related to microscopic biology and nanoscale technologies. This work thus sets the basis for truly nanoscale investigations of Taylor dispersion.

\begin{acknowledgments}
\noindent The authors gratefully acknowledge David Lacoste, Andreas Engel, Arthur Alexandre, Thomas Gu\'erin and David Dean for enlightening discussions. Patrick Tabeling and Fabrice Monti are likewise thanked for helpful advice related to TIRFM. The authors also benefited from the financial support of CNRS, ESPCI Paris, the Agence Nationale de la Recherche (ANR) under the ENCORE (ANR-15-CE06-005) and CoPinS (ANR-19-CE06-0021) grants, and of the Institut Pierre-Gilles de Gennes (Equipex ANR-10-EQPX-34 and Labex ANR-10-LABX- 31), PSL Research Uniersity (Idex ANR-10-IDEX-0001-02).
\end{acknowledgments}


\clearpage

\appendix

\renewcommand\thefigure{A.\arabic{figure}}    
\setcounter{figure}{0} 

\section{\label{sec:Methods}{Methods}}

\noindent The experiments employed pressure-driven flows (Fluigent MFCS-4C pressure controller) in microchannels with a rectangular section using the liquids noted above. Microfluidic chips were fabricated by soft lithography of poly(dimethyl siloxane) (Dow Chemical, Sylgard 184) on a plasma-cleaned glass coverslip with 145 $\mu$m thickness constituting the bottom surface. The liquids used were ultra-pure water (18.2 M$\Omega$\ cm, MilliQ) and water-glycerol mixtures with Newtonian viscosities of $\eta = 1, 2.1\mathrm{\ and\ }7.6$ mPa\ s, measured with a Couette-cell rheometer (Anton Paar MCR 302) up to $\dot{\gamma} = 1000\ \mathrm{s}^{-1}$. The carboxylate-modified fluorescent nanoparticles used were 55 nm-radius (Invitrogen F8803, Thermofisher) and 100 nm-radius (Invitrogen F8888 Thermofisher) latex microspheres used without further modification besides dilution by a factor of $10^3$ using ultra-pure water. 

TIRFM measurements, as in~\cite{Hoffman2011, Li2015}, were realized by illuminating the near-surface shear flow with a laser source (Coherent Sapphire, wavelength $\lambda=488$ nm, power 150 mW) focused off the central axis of, and on the back focal plane of a $100\times$ microscope objective with a large numerical aperture ($\mathrm{NA}=1.46$, Leica HCX PL APO). Thus incident angles $\theta$ larger than the critical angle, $\theta_\mathrm{c} =\arcsin(n_\mathrm{l}/n_\mathrm{g})$, were reached enabling total reflection at the glass/liquid interface. Here, $n_\mathrm{g}=1.518$ is the refractive index of the glass coverslip, $n_\mathrm{f} = \{1.33,\ 1.36,\ 1.40\}$ is the refractive index of the fluids for the water and glycerol mixtures, respectively, and $\theta$ the angle of incidence of the laser; the refractive indices of the three liquids were measured using a refractometer (Atago PAL-RI). The setups give rise to exponential decay lengths $\Pi=\lambda/(4\pi)\left(n_\mathrm{g}^2\sin^2\theta-n_\mathrm{f}^2\right)^{-1/2}$ with the angle measured \emph{in situ} as in ref.~\cite{Guyard2021}. The penetration depth was thus $\Pi\approx 100$ nm. The images of $528\times512$ pixels (px), with 22.9 px/$\mu$m, are recorded in 16-bit format (Andor Neo sCMOS) with a frame rate of 400 Hz for a duration of 5 s. For each set of parameters, five videos of 2000 frames were recorded. After a centroid detection, the intensity profile was fitted by a radially-symmetric Gaussian model for each frame as shown in Fig.~\ref{fig:Fig1}b). Thus the $x$ and $y$ coordinates give the particle position in the plane parallel to the glass/water interface.

\section{\label{sec:SIDmeanV}Intensity distribution and mean velocity profiles}

\noindent This section provides additional information about how the observed signal intensity distributions (SIDs after Zheng and coworkers \cite{Zheng2018}, denoted $\mathcal{P}_\mathrm{SID}$) and the corresponding velocity profiles can be quantitatively described simultaneously. As also described by Li and coworkers \cite{Li2015}, fluorescent nanoparticles detected display a range of intensities affected by several factors, the most important ones being: \emph{i}) electrostatic interactions which determine the probability that a particle of a given radius is found at a certain distance from the wall according to a Boltzmann distribution; \emph{ii}) particle size distribution; and \emph{iii}) the optical setup which, given the position and size of the particle, finally determines its intensity. We now discuss each of these elements in detail.  

\vspace{2mm}

\noindent{\emph{i}) The glass surface exerts an electrostatic repulsion on the particles according to the fact that both surfaces are negatively charged; the details of such a repulsion are understood within the Derjaguin-Landau-Verwey-Overbeek (DLVO) framework \cite{Derjaguin1940,Verwey1947}. This electrostatic interaction potential, $\phi_\mathrm{el}$, describing the electric double-layer repulsion between a particle with radius $R$ and a flat wall~\cite{Prieve1999} is given by:
\begin{equation}
	\phi_{\text{el}}\left(z\right) = 16\epsilon R\left(\frac{k_{\mathrm{B}}\Theta}{e}\right)^{2}\tanh\left(\frac{e\psi_{\mathrm{p}}}{4k_{\mathrm{B}}\Theta}\right)\tanh\left(\frac{e\psi_{\mathrm{w}}}{4k_{\mathrm{B}}\Theta}\right)\exp\left(-\frac{z-R}{l_{\mathrm{D}}}\right)\ .
	\label{DLVO}
\end{equation}
Furthermore $z$, $\epsilon$, $e$, $\psi_{\mathrm{p}}$, $\psi_{\mathrm{w}}$ and $l_{\mathrm{D}}$ are respectively the position of the center of the particle, liquid permittivity, elementary charge, particle and wall electrostatic potentials and the Debye length. This interaction determines the particle concentration $C$ at thermal equilibrium through the Boltzmann distribution 
\begin{equation}
C(z)\propto\exp\left(-\frac{\phi_\mathrm{el}\left(z\right)}{k_{\mathrm{B}}\Theta}\right)\ .
\label{Boltzmann}
\end{equation}
As already observed in TIRFM experiments, the van der Waals interaction can be neglected for pure water~\cite{Zheng2018,Li2015}. Consequently, the typical distance between the bottom surface (located at $z=0$) and the particles is mainly determined by the Debye length. 

\vspace{2mm}

\noindent\emph{ii}) All the particles do not have the same radius $R$. The radius distribution is described by a Gaussian probability function 
\begin{equation}
	\mathcal{P}_R\left(R\right)=\frac{1}{\sqrt{2\pi \sigma_R^{\,2}}}\exp\left(-\frac{\left(R-a\right)^2}{2\sigma_R^{\,2}}\right)\ ,
	\label{PofR}
\end{equation}
where $a$ is the mean radius and $\sigma_R$ the standard deviation.

\vspace{2mm}

\noindent\emph{iii}) The fluorescence intensity, $I$, of an individual particle is determined by the optical parameters of the TIRFM setup and the particle's size, with $I\propto R^3$. The evanescent wave has a penetration depth $\Pi$ characterizing the exponential decrease of excitation. The observed fluorescence intensity is also sensitive to the finite depth of field, $d_\mathrm{f}$, of the microscope objective. In our experiments, the depth of field has a value of 415 nm, meaning that if particles are not located on the focal plane at $z_\mathrm{f}$ (typically 400-500 nm from the glass-liquid interface), they will be detected with a relatively low intensity. Putting these elements together, the observed fluorescence intensity for an individual particle is predicted~\cite{Zheng2018} as
\begin{equation}
	\frac{I\left(R,z\right)}{I_0}=\left(\frac{R}{a}\right)^3\exp\left(-\frac{z-a}{\Pi}\right)\left[1+\left(\frac{z-a-z_\mathrm{f}}{d_\mathrm{f}}\right)^2\right]^{-1}\ ,
	\label{EvanDoF}
\end{equation}
where $I_0$ is the intensity for a particle with radius $R=a$ located at the bottom surface $z=a$ and the focal plane at $z_\mathrm{f}=a$.

Using a home-made \textsc{Matlab} interface, we combine Eqs.~\eqref{DLVO}-\eqref{EvanDoF} to generate numerical SIDs. Practically, we determine the fraction of particles having an altitude $z$ and a radius $R$ given by the weight $W(z,R)=C(z)\mathcal{P}_R(R)$, and compute the associated intensity given by Eq.~\eqref{EvanDoF}. This procedure gives a list of weighted intensities forming the blue line shown in Fig.~\ref{fig:SIDMeanVelocity}(a) using a DLVO prefactor $16 a \epsilon(k_\mathrm{B}\Theta/e)^2\tanh(e\psi_\mathrm{p}/(4k_\mathrm{B}\Theta))$ $\tanh(e\psi_\mathrm{w}/(4k_\mathrm{B}\Theta)) = 1.4\times 10^{-21}$~J, $l_\mathrm{D} = 60$\,nm, $\sigma_R = 5.5$\ nm, $a=55$ nm and the optical parameters as described above, along with the experimental histogram (red).  
\begin{figure}\centering
	\includegraphics[scale=1.0,trim=0 0 0 0,clip=true]{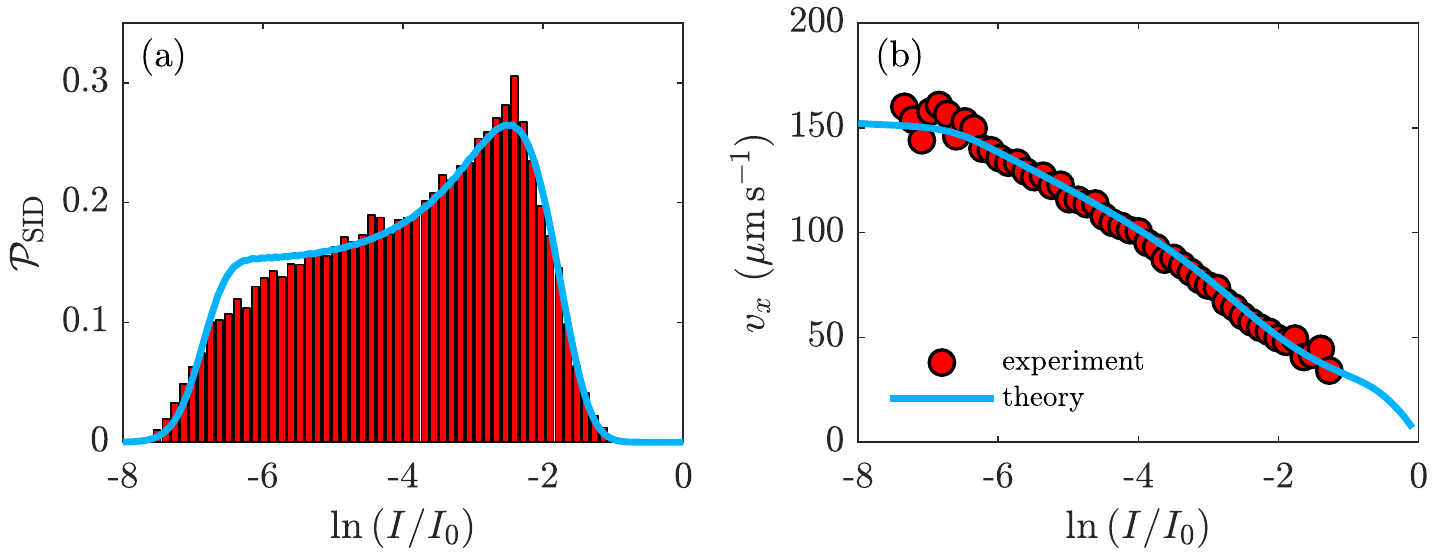}
	\caption{\textit{(a) Comparison between experimental and theoretical signal intensity distributions (SIDs). (b) Comparison between experimental and theoretical streamwise mean velocity profiles. The experimental data is for 55 nm-radius particles for a pressure drop of 30 mbar across the microchannel.}}
\label{fig:SIDMeanVelocity}
\end{figure}

In addition, we also quantitatively describe the mean streamwise velocity profile, which, as noted before \cite{Li2015}, is not perfectly linear when using the apparent altitude $z=a+\Pi\ln\left(I_0/I\right)$ defined in the main text. We assume that a particle located at an altitude $z$ has a mean streamwise velocity $v_x$ given by $v_x\left(z\right)=f_\mathrm{B}\dot{\gamma}z$, where $\dot{\gamma}$ is the shear rate and $f_\mathrm{B}$ the ``Brenner factor" \cite{Goldman1967}. This factor provides the hydrodynamic correction induced by the finite size of the spherical particle, when the latter is advected by a linear shear flow near a wall. For large $z/R$, the Brenner factor can be expressed as $f_\mathrm{B}\simeq 1-\left(5/16\right)\left(z/R\right)^{-3}$	. For 55 nm-radius particles typically located at distances larger than 200 nm due to electrostatic repulsion, the deviation from the linear velocity profile is less than $1\%$. 

Using the proposed particle velocity profile in conjunction with the intensity-altitude-probability relations (Eqs.~\eqref{DLVO}-\eqref{EvanDoF}), we follow Zheng \textit{et al.} \cite{Zheng2018} and predict the particle's mean streamwise velocity as a function of $\ln(I_0/I)$. Such a prediction is made with $\dot{\gamma}$ adjusted simultaneously to the physical and optical parameters of Eqs.~\eqref{DLVO}-\eqref{EvanDoF}. The result is shown together with the experimental results in Fig.~\ref{fig:SIDMeanVelocity}(b), showing good agreement and capturing the main nonlinear features of the experimental data. The shear-rate values obtained with this SID method are approximately $15\%$ smaller than the ones directly obtained using a linear regression of the velocity profiles of Fig.~1(c) using the apparent altitude. This discrepancy is mainly due to the particle polydispersity and to the finite depth of field of the microscope objective, and since it is only a constant factor (verified) across all experiments it does not change the main conclusions of the article.

\section{\label{sec:ViscoDiff}Medium viscosity and particle diffusion}

\noindent In Fig.~1(c), we show the streamwise velocity profiles for 55 nm-radius particles in a water flow obtained by total internal reflection fluorescence microscopy (TIRFM). In the corresponding inset, we show the associated shear rate $\dot{\gamma}$ (obtained from a linear regression on a given velocity profile) as a function of the pressure drop $\Delta P$ across the channel. Similar measurements were done for the 100 nm-radius particles in water and water-glycerol mixtures presented in the main article, see Figs.~3 and 4. Corresponding to these shear rate measurements, we can compute the stress, $\Sigma=h\Delta P/(2L)$, from the pressure drop across the rectangular channel using a geometric prefactor (height $h=18$ $\mu$m, width $w=180$ $\mu$m, length $L=8.8$ cm). The stress $\Sigma$ is plotted as a function of the shear rate $\dot{\gamma}$ in Fig.~\ref{fig:AllViscoDiff}(a) and compared with bulk rheology measurements carried out in a Couette cell (see Methods).
\begin{figure}\centering
	\includegraphics[scale=1.0,trim=0 0 0 0,clip=true]{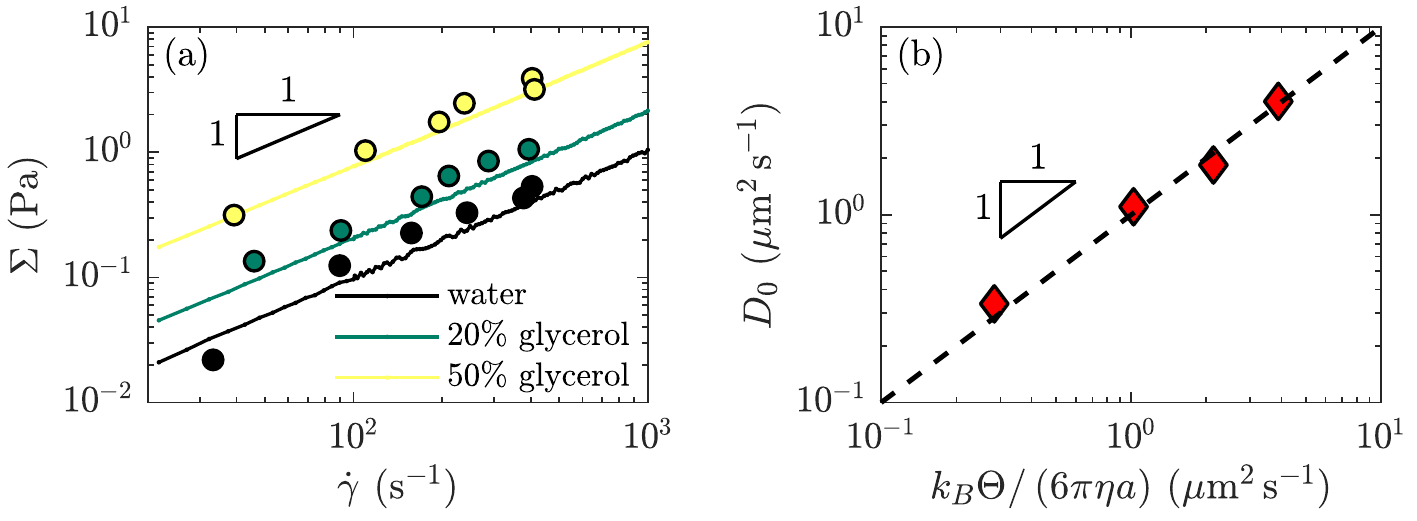}
	\caption{\textit{(a) Comparison for stresses $\Sigma$ versus shear rate $\dot{\gamma}$ between TIRFM (circles) and rheology measurements (lines) for 100 nm-radius particles, see Fig.~\ref{fig:Fig2}(b) for $a = 55$ nm particles in water. For TIRFM, the stress is calculated as $\Sigma=h\Delta P/(2L)$ where $\Delta P$ is the pressure drop in the channel, $h$ and $L$ are the channel height and length. (b) Bulk diffusion coefficient $D_0$, measured from the plateau values of local transverse mean square displacements for two particle sizes and three viscosities, versus the theoretical values calculated using, $k_\mathrm{B}\Theta$, the thermal energy at room temperature, and the viscosity measured with a rheometer. The black dashed line shows the linear relation with unit prefactor.}}
\label{fig:AllViscoDiff}
\end{figure}
First, the resulting linear power laws show that all the solutions remain Newtonian for shear rates up to 1000 s$^{-1}$. Second, the viscosity defined as $\eta=\Sigma/\dot{\gamma}$ is consequently constant for a given solution, and it increases with the glycerol proportion. The results show a good agreement with the rheology measurements, validating both the shear rate and viscosity values obtained by TIRFM.

The viscosity values obtained further allow us to verify that the bulk diffusion coefficients measured with TIRFM are consistent with the Stokes-Einstein relation. As shown in Fig.~4(a) in the main article, the bulk diffusion coefficient $D_0$ is obtained from the plateau value of the local transverse diffusion coefficient $D_y$, calculated from the transverse mean-square displacement $\sigma_{\Delta y}^2$, through: $D_y=\sigma_{\Delta y}^2/(2\tau)$, where $\tau$ is the lag time. In Fig.~\ref{fig:AllViscoDiff}(b) is shown a comparison between the experimental results and the prediction given by the Stokes-Einstein relation \cite{Einstein1905}: $D_0=k_{\mathrm{B}}\Theta/(6\pi\eta a)$, after having used the independently measured viscosity. The good agreement validates the statistical method to obtain the bulk diffusion coefficient.

\end{document}